\documentclass[a4paper,11pt]{article}

\pdfoutput=1

\usepackage{jheppub,fancyhdr,epstopdf,color,amsmath,cases,slashed,hyperref,subfigure}
\usepackage{amsfonts}
\usepackage{amssymb}
\usepackage{graphicx,graphics,epsfig}
\usepackage{geometry}
\usepackage[english]{babel}
\usepackage{mathtools}
\usepackage[utf8]{inputenc}
\usepackage[T1]{fontenc}
 \usepackage{float}
\usepackage{hyperref}
\usepackage{ulem}
\usepackage{caption}
\usepackage{caption}
\usepackage{subfigure}  
\usepackage{mathrsfs}  
\usepackage{bbold}
\usepackage{comment}

\def\noi{\noindent}
\def\non{\nonumber}

\newcommand{\msbar}{\overline{\rm MS}}

\def\l{\lambda}

\def\g{\gamma}

\newcommand{\Lag}{\mathcal{L}}

\def\ba{\begin{array}}
\def\ea{\end{array}}
\def\bea{\begin{eqnarray}}
\def\eea{\end{eqnarray}}
\newcommand{\Omegah}{\Omega h^{2}}

\newcommand{\Asla}{\not{\hbox{\kern-3.5pt $A$}}}
\newcommand{\Gsla}{\not{\hbox{\kern-3.5pt $G$}}}
\newcommand{\Wsla}{\not{\hbox{\kern-3.5pt $W$}}}
\newcommand{\Zsla}{\not{\hbox{\kern-3.5pt $Z$}}}

\newcommand{\Dslash}{\not{\hbox{\kern-4pt $D$}}}
\newcommand{\pslash}{\not{\hbox{\kern-2.3pt $p$}}}

\def\lsim{\;\raise0.3ex\hbox{$<$\kern-0.75em\raise-1.1ex\hbox{$\sim$}}\;}
\def\gsim{\;\raise0.3ex\hbox{$>$\kern-0.75em\raise-1.1ex\hbox{$\sim$}}\;}

\def\l{\lambda}

\def\g{\gamma}

\def\ba{\begin{array}}
\def\ea{\end{array}}
\def\bea{\begin{eqnarray}}
\def\eea{\end{eqnarray}}

\def\lsim{\;\raise0.3ex\hbox{$<$\kern-0.75em\raise-1.1ex\hbox{$\sim$}}\;}
\def\gsim{\;\raise0.3ex\hbox{$>$\kern-0.75em\raise-1.1ex\hbox{$\sim$}}\;}

\newcommand{\beqn}{\begin{eqnarray}}
\newcommand{\eeqn}{\end{eqnarray}}

\title{Relic density of Dark Matter in the Inert Doublet Model beyond Leading Order. I) The  Heavy Mass Case}

\preprint{LAPTH-035/19, IPPP/19/52, NCTS-PH/1906}

\author[a]{Shankha Banerjee}
\author[b]{\!\!, Fawzi Boudjema}
\author[c]{\!\!, Nabarun Chakrabarty}
\author[d]{\!\!,  Guillaume Chalons}
\author[e]{\!\!,  Hao Sun}

\affiliation[a]{Institute for Particle Physics Phenomenology, Department of Physics, Durham University, Durham DH1 3LE, United Kingdom}
\affiliation[b]{LAPTh, Universit\'e Savoie Mont Blanc, CNRS, BP~110, F-74941 Annecy-le-Vieux, France}
\affiliation[c]{Physics Division, National Center for Theoretical Sciences, Hsinchu, Taiwan 30013, R.O.C.}
\affiliation[d]{Laboratoire de Physique Subatomique et de Cosmologie, Université Grenoble-Alpes, CNRS/IN2P3, 53 Avenue des Martyrs, F-38026 Grenoble, France}
\affiliation[e]{Institute of Theoretical Physics, School of Physics, Dalian University of Technology, Dalian 116024, People’s Republic of China}

\emailAdd{shankha.banerjee@durham.ac.uk}
\emailAdd{boudjema@lapth.cnrs.fr}
\emailAdd{nchakrabarty@cts.nthu.edu.tw}
\emailAdd{chalons@lpsc.in2p3.fr}
\emailAdd{haosun@dlut.edu.cn}

 \abstract{A full renormalisation of the Inert Doublet Model (IDM) is presented and exploited for a precise calculation of the relic density of Dark Matter (DM) at one-loop. In this first paper, we study the case of a DM candidate with $m_{\rm DM} \sim 500$ GeV. In this regime, the co-annihilation channels are important. We therefore compute, for a wide range of relative velocities, the full next-to-leading order electroweak corrections to 7 annihilation/co-annihilation processes that contribute $\sim$ 70\% to the relic density of DM. These corrected cross-sections are interfaced with {\tt micrOMEGAs} to obtain the one-loop correction to the freeze-out relic density. Due to the accurate measurement of this observable, the one-loop corrections are relevant. We discuss the one-loop renormalisation scheme dependence and point out the influence, at one-loop, of a parameter that solely describes the scattering in the dark sector. A tree-level computation of the relic density is not sensitive to this parameter.}
 
\begin{document}
 
\date\today

\maketitle
\flushbottom

\section{Introduction}
The inert double model (IDM) consists in adding a scalar doublet, $\Phi_2$, to the Standard Model of particle physics (SM)~\cite{Deshpande:1977rw}. Endowing this most simple extension of the SM with an unbroken $\mathbb{Z}_2$ symmetry where $\Phi_2$ is odd while other fields of the SM are even, guarantees the stability of the lightest inert particle, thus providing a possible dark matter (DM) candidate~\cite{Barbieri:2006dq}. The new scalars of this additional doublet couple to the  Higgs and the gauge bosons but not to the fermions in the SM. This model provides a nice link between the Higgs sector with the source of electroweak symmetry breaking and DM~\cite{Hambye:2007vf}. The model has received a lot of attention, primarily for DM studies but also for collider observables, see~\cite{Cao:2007rm,Hambye:2009pw, Goudelis:2013uca, Arhrib:2013ela, Queiroz:2015utg,Belyaev:2016lok,Arcadi:2019lka} for reviews and updates. The majority of these analyses were performed at leading order. A few exceptions where one-loop effects are considered include {\it i)} the computation of the tri-linear self-coupling of the SM-like Higgs boson~\cite{Kanemura:2002vm, Senaha:2018xek, Braathen:2019pxr, Arhrib:2015hoa}, {\it ii)} one-loop corrections to the Higgs effective potential~\cite{Hambye:2007vf,Ferreira:2015pfi}, the running of the Higgs/scalar masses and the running of the scalar parameters~\cite{Goudelis:2013uca} (see also~\cite{Ferreira:2009jb}), {\it iii)}  one-loop induced cross-sections relevant for direct detection~\cite{Klasen:2013btp} and {\it iv)} induced one-loop effects for photon production: Higgs decay to a photon pair~\cite{Arhrib:2012ia} and DM annihilation to photons~\cite{Gustafsson:2007pc,Garcia-Cely:2015khw}. Yet, the relic density of DM as extracted by PLANCK~\cite{Ade:2015xua} is a precision measurement at the percent level that calls for an equally precise theoretical prediction. In particular, the perturbative DM annihilation cross-sections that drive the amount of relic density must be evaluated beyond tree-level. This has not been performed in the IDM case despite the fact that the relic density sets the most stringent constraint on the IDM. This is somehow understandable since this task requires a coherent full renormalisation of the model, the evaluation of many processes at one-loop order and the inclusion of these processes for the evaluation of the relic density. It is the purpose of this paper to present such a programme and to present the first results for one-loop corrected processes and how they affect the value of the the freeze-out relic density in the IDM.  

The IDM model is of course subject to various experimental constraints that leave two viable scenarios: one where the DM mass is about $M_W$ and the other where the mass is above $500$ GeV. We will, in this first paper, be interested in a scenario with $M_{\rm{DM}} \sim 500$ GeV. Non perturbative effects, such as the importance of the electroweak Sommerfeld effects~\cite{Hisano:2002fk, Hisano:2003ec, Hisano:2004ds, Hambye:2009pw, Biondini:2017ufr}, occur for very high masses above the TeV and will not be treated here.

The plan of the paper is as follows. In the next section~\ref{sec:tree-level} we will first outline the model and underline its parameters. The renormalisation procedure is exposed in section~\ref{sec:renormalisation}. Section~\ref{sec:benchmarks} takes into account the various constraints on the model and motivates us in setting up our benchmark scenario. Section~\ref{sec:loopresults} presents our findings for the full next-to-leading order corrections that, at tree-level, contributes more than 5\% of the relic density contribution. We will find that because of co-annihilation we will need to consider 7 processes. Section~\ref{sec:relic-one-loop} will translate these improved predictions into a corrected value of the relic density by interfacing our cross-sections with {\tt micrOMEGAs}. Finally we conclude our findings in section~\ref{sec:conclusions}. The appendices are relegated to section~\ref{sec:appendix}.

\section{The Inert Doublet Model at the classical level}
\label{sec:tree-level}

To the Higgs doublet $\Phi_1$ of the SM, a doublet $\Phi_2$ is added. An unbroken $\mathbb{Z}_2$ symmetry is imposed under which $\Phi_2$ is odd while all other fields (of the SM) are even. The immediate consequence is that $\Phi_2$ cannot couple to fermions to any order (in perturbation theory) and guarantees the stability of the lightest inert particle, thus providing a possible dark matter candidate. The
Lagrangian of the IDM can be written as,

\begin{equation}
\Lag_{IDM} = \Lag_{SM} +  (D^\mu \Phi_2)^\dagger D_\mu \Phi_2 +
\mathscr{V}_{IDM}(\Phi_1,\Phi_2), 
\end{equation}\noi

where $\Lag_{SM}$ is the SM Lagrangian whereas the scalar potential is given by

\begin{eqnarray}
\label{IDMpot}
 \mathscr{V}_{IDM}(\Phi_1,\Phi_2) &=& \mu_1^2 |\Phi_1|^2 + \mu_2^2 |\Phi_2|^2 +
\l_1 |\Phi_1|^4 + \l_2 |\Phi_2|^4  \non \\
& & + \l_3 |\Phi_1|^2 |\Phi_2|^2 +
\l_4 (\Phi_2^\dagger \Phi_1)(\Phi_1^\dagger \Phi_2) +
\left(\frac{\l_5}{2}(\Phi_1^\dagger \Phi_2)^2 + \mbox{h.c} \right).
\end{eqnarray}\noi

In this equation, $\mu_i$ and  $\l_i$ are real. Since the unbroken $\mathbb{Z}_2$ symmetry prevents the presence of tadpole terms for $\Phi_2$ (and therefore no vacuum expectation value from $\Phi_2$) and mixing with $\Phi_1$, we can directly parametrise the doublets in terms of the physical scalars,

\begin{equation}
 \Phi_1 = \begin{pmatrix}
           G^+ \\
	\frac{1}{\sqrt{2}}\left(v + h + i G\right)
          \end{pmatrix}~\mbox{and}~
\Phi_2 = \begin{pmatrix}
          H^+ \\
    \frac{1}{\sqrt{2}}\left(H + i A \right)
         \end{pmatrix},
\end{equation}\noi

where $v$ is the SM vacuum expectation value (vev) with $v \simeq 246$ GeV, defined from the measurement of the $W$ ($M_W$) and $Z$ ($M_Z$) masses. We have

\begin{equation}
\label{eq:def_sw2}
s_W^2 \equiv \sin^2 \theta_W = 1 - \frac{M_W^2}{M_Z^2}, \quad 
M_W = \frac{1}{2}\frac{e}{s_W} v,
\end{equation}\noi

\noi $e$ is the electromagnetic coupling (the $SU(2)$ gauge coupling $g$ is then $g=e/s_W$), $h$ is the SM 125 GeV Higgs boson, $G,G^\pm$ are the Goldstone bosons, $H,A$ are the new neutral physical scalars~\footnote{Since these additional scalars do not couple to the fermions (of the SM), we can not assign them definite CP numbers. By an abuse of language, we will call $A$ the pseudo-scalar.} and $H^\pm$ is the charged physical scalar. $H$ and $A$ are the possible DM candidates. These scalars have gauge couplings to the SM gauge bosons, controlled by the SM gauge coupling. For example, for the tri-linear couplings we have 

\begin{align}
\label{eq:gauge_couplings}
(H^+H^-\gamma, \; H^+ H^-Z, \; H H^\pm W^\mp, \; iA H^\pm W^\mp, \; iA H Z) = \nonumber \\
i\frac{g}{2}(2 s_W, \; c_{2W}/c_W, \; \mp 1, \; -1, \; -1/c_W). 
\end{align}\noi

We must note that quartic couplings of the type $HHW^+W^-$ are also present. Annihilation of DM to vector bosons proceeds, in part, through these gauge interactions and in part through the scalar potential coupling to which we now turn our attention for more details.

\subsection{Minimisation of the potential}
Minimisation of the potential amounts to vanishing tadpoles for $\Phi_1$ leading to the constraint

\begin{equation}
\label{tadeq}
\frac{T}{v} = \mu_1^2 + \l_1 v^2 \equiv 0.
\end{equation}\noi

There is no corresponding tadpole term for $\Phi_2$ because of the unbroken $\mathbb{Z}_2$ symmetry. The no-tadpole condition will be maintained at all orders.

\subsection{Mass spectrum and scalar self-interactions}
By collecting the bilinear terms in the physical scalar fields of $\mathscr{V}_{IDM}$, we get the mass spectrum of the scalar sector of the IDM

\begin{align}
\label{hmass}
M_h^2 &= \frac{T}{v} +  2 \l_1 v^2,\\
\label{Hpmass}
M_{H^\pm}^2 &= \mu_2^2 + \l_3 \frac{v^2}{2},\\
 \label{Hmass}
M_H^2 &= \mu_2^2 + \l_L \frac{v^2}{2}=  M_{H^\pm}^2 +
\left(\l_4+\l_5 \right)  \frac{v^2}{2},\\
 \label{Amass}
M_A^2 &= \mu_2^2 + \l_A \frac{v^2}{2}= M_{H^\pm}^2 +
\left(\l_4-\l_5\right) \frac{v^2}{2}= M_H^2 - \l_5 v^2,
\end{align}\noi 
where 
\beqn
\l_{L/A} = \l_3+\l_4\pm \l_5.
\eeqn

In the following, we will consider $H$ as the possible DM candidate. The underlying reason is that both $H$ and $A$ can be treated on equal footing. Choosing $A$ as the DM candidate would simply correspond to a flip in the sign $\l_5 \to -\l_5$ without changing the phenomenology. The reason is the following (we borrow arguments from \cite{Ilnicka:2015jba}): taking $H$ as the DM candidate and thus $M_H < M_A, M_{H^\pm}$, from Eqs.~\ref{Hmass} and~\ref{Amass} we obtain

\begin{equation}
 \l_4 + \l_5 < 0 \quad \mbox{and} \quad \l_5 < 0.
\end{equation}\noi

The converse situation, with $M_A < M_H, M_{H^\pm}$, corresponds to,

\begin{equation}
 \l_4 - \l_5 < 0 \quad \mbox{and} \quad \l_5 > 0.
\end{equation}\noi

For $\l_5=0$, $H$ and $A$ are mass degenerate. All portal triple and quartic couplings of the SM-like Higgs $h$ to $H/A$ are proportional to $\l_{L/A} $. Indeed, we can write, at tree-level, the $hHH$ and $hAA$ coupling as 

\beqn
\lambda_{hHH}=\lambda_L v, \quad \lambda_{hAA}=\lambda_A v. 
\eeqn

Thus considering one or the other scalars as the DM candidate amounts to switching

\begin{equation}
\l_5 \longleftrightarrow - \l_5, \l_L \longleftrightarrow \l_A.
\end{equation}

In the same vein, we can write 

\beqn
\lambda_{h H^+ H^-} = \l_3 v.
\eeqn

The quartic couplings between the SM Higgs and the new scalar are set by $\lambda_{3,L,A}$,

\beqn
\lambda_{hhHH,hhAA,hhH^+H^-}=\l_L,\l_A,\l_3.
\eeqn
On the other hand, $\lambda_2$ controls all the quartic couplings solely within the dark sector ($HHHH$, $HHAA$, $HHH^+H^-$, $AAAA$, $AAH^+H^-$ and $H^+H^-H^+H^-$).

\subsection{Counting parameters}
In order to survey the IDM parameter space it is important to count the number of independent parameters in the scalar sector. Setting aside the tadpole condition and the 125 GeV (SM) Higgs mass, the IDM requires 5 extra  parameters,

\begin{equation}
 \left(\mu_2,\l_2,\l_3,\l_4,\l_5\right).
\end{equation}\noi
\par\noi

It is interesting to trade 3 of the above parameters of the scalar sector for the physical masses of the new scalars through Eqs.~(\ref{Hpmass}-\ref{Amass}). This will be important for the renormalisation programme when we adopt an on-shell scheme~\footnote{Note that $s_W$ was also defined in terms of the $W$ and $Z$ masses, Eq.~\ref{eq:def_sw2}.}. The model can therefore be defined through the following two possible trade-offs,

\begin{equation}
\label{inputparam}
\left(\mu_2,\l_3,\l_4,\l_5; \l_2\right) \to \left(
M_H,M_A,M_{H^\pm},\l_{L/A};\l_2\right), 
\end{equation}\noi

or equivalently

\begin{equation}
 \left(\mu_2,\l_3,\l_4,\l_5; \l_2\right) \to \left(
M_H,M_A,M_{H^\pm}, \mu_2; \l_2\right).
\end{equation}
\noi We set $\l_2$ apart as it describes couplings solely between the additional scalars and  not involving the SM Higgs. At tree-level for example and for $2\to 2$ annihilation processes, $\l_2$  is irrelevant. This would mean that at one-loop order, for annihilation processes, a renormalisation for $\l_2$ is not necessary. However, $\l_4$ and $\l_5$ can be reconstructed from a combination of the additional scalar masses
\begin{align}
\label{l4eq}
 \l_4 &= \frac{1}{v^2}\left(M_H^2 + M_A^2 - 2 M_{H^\pm}^2\right), \\
\label{l5eq}
 \l_5 &= \frac{1}{v^2}\left(M_H^2 - M_A^2\right).
\end{align}\noi 

\noi The extraction of $\l_3$ not only requires a knowledge of at least one scalar mass but also either a value of $\lambda_L$ (or equivalently the $hHH$ coupling) or the mass parameter $\mu_2$, to wit 

\begin{align}
\label{l3eq}
 \l_3 &= \frac{2}{v^2}\left(M_{H^\pm}^2 - \mu_2^2\right)= \frac{2}{v^2}\left(M_{H^\pm}^2 - M_H^2\right)+ \l_L.
\end{align}

\section{Renormalisation of the IDM}
\label{sec:renormalisation}
The presence of the $\mathbb{Z}_2$ symmetry tremendously eases the renormalisation of the IDM. As a result of this symmetry there is no mixing, at any order, between the SM fields and the extra fields introduced by the IDM. The tadpole  condition only applies to the SM part. The SM part, including the SM Higgs (and the Goldstone bosons), are renormalised, independently and exactly as in the SM. We therefore follow an On-Shell (OS) scheme whose details can be found in Ref.~\cite{Belanger:2003sd}. We will pursue the OS approach for all three extra physical scalar fields, $H,A,H^\pm$. We will therefore use the physical masses of these fields as input parameters instead of the parameters of the scalar potential. Nonetheless, there remain 2 parameters which we still need to define. As stated earlier, of all the parameters in the IDM, only $\l_2$ connects the extra fields. Its renormalisation is not needed for one-loop annihilation processes to SM particles. However, one parameter (either $\mu_2$ or $\lambda_{L,A}$ or a combination of these) is still needed to fully define all the couplings between $H,A,H^\pm$ and the SM Higgs and Goldstones, see Eq.~\ref{inputparam}. For example $\l_{L/A}$ has, at tree-level, a simple physical interpretation as the portal coupling
$hHH/hAA$. 

To carry the renormalisation programme and define the counterterms, shifts are  introduced for the Lagrangian parameters and the fields. All bare quantities ($X_0$), particularly in Eq.~\ref{IDMpot}, are  decomposed into renormalised quantities ($X$) and counterterms ($\delta X$) as

\begin{equation}
 X_0 \to X + \delta X, \quad X=\mu_2,\l_2, \l_3,\l_4,\l_5,
\end{equation}\noi 

for the parameters~\footnote{This procedure is also applied to the SM sector including the $\mu_1$ and $\lambda_1$ terms of potential. For the latter, the tadpole condition is imposed at one-loop; see~\cite{Belanger:2003sd}} and
\begin{equation}
 \phi_0 \to \phi + \frac{1}{2}\delta Z_\phi, \quad \phi=(h,H,A,H^\pm),
\end{equation}\noi 
for the fields.

The OS conditions on the physical scalars  require that their masses  are defined as pole masses of the renormalised one-loop propagator and that the residue at the pole be unity. With $\Sigma_{\phi\phi}(p^2)$ being the scalar two-point function with momentum $p$ we have ($\phi = h,H,A,H^\pm$),
\begin{align}
\label{ct_deltam}
 \delta M_\phi^2 &= \Sigma_{\phi\phi}(M_\phi^2) \\
 \label{ct_deltaZ}
 \delta Z_\phi &= - \left.\frac{\partial \Sigma_{\phi\phi}(p^2)}{\partial
p^2}\right|_{p^2 = M_\phi^2}.
\end{align}\noi 

$\delta M_H, \delta M_A$ and $\delta M_{H^\pm}$ directly give OS definitions for $\l_4$ and $\l_5$ through Eqs.~\ref{l4eq} and~\ref{l5eq}. With only 3 physical masses, we can not reconstruct all the 5 counterterms from two-point functions.  We therefore revert to couplings between the Higgses. Remembering that $\l_L$ measures the $hHH$ coupling, we could extract a counterterm for $\delta \l_L$ from a measurement of the $hHH$ coupling. We could have also chosen the $hAA$ or $hH^+ H^-$ couplings. Sticking with $\lambda_L$,  when $M_h > 2 M_H$, the invisible width of the Higgs $\Gamma(h \to HH)$ is the observable of choice especially because no infra-red divergence affects this observable. This observable will therefore be set as input, which is equivalent to stating that the observable receives no correction. We denote the amplitude for $h \to HH$ as ${\cal{A}}(h \to HH) \equiv {\cal{A}}_{hHH}$. We express this amplitude at tree-level as 

\beqn
{\cal{A}}_{hHH}^0=-\l_L v,
\eeqn

and the full one-loop renormalised amplitude for $h(p^2) \to H(p_1^2) H(p_2^2)$ as 
\beqn
{\cal{A}}_{hHH}^{\rm ren}(p^2,p_1^2,p_2^2)=- \l_L v \left(\frac{\delta
\l_L}{\l_L} + \frac{\delta v}{v}+ \frac{1}{2}\delta Z_h + \delta Z_H \right) + {\cal{A}}^{\rm 1PI}_{HHh}(p^2, p_1^2,p_2^2),
\eeqn
where ${\cal{A}}^{\rm 1PI}_{HHh}(p^2,p_1^2,p_2^2)$ is the full one-loop 1 particle irreducible vertex. When the threshold is open, we set $p^2=M_h^2$ and $p_1^2=p_2^2=M_H^2$ defining a gauge invariant OS counterterm for $\l_L$ as 

\begin{equation}
 \frac{\delta^{{\rm OS}} \l_L}{\l_L} = 
\frac{{\cal{A}}^{\rm 1PI}_{HHh}(m_h^2,M_H^2,M_H^2)}{\l_L v}
-\frac{\delta v}{v} -\frac{1}{2}\delta Z_h -\delta
Z_H.
\end{equation}\noi 
Another gauge invariant but scale dependent scheme is to use a $\overline{{\rm MS}}$ definition where only the (mass independent term) ultraviolet  divergent part is kept

\begin{equation}
 \frac{\delta^{\overline{{\rm MS}}} \l_L}{\l_L} = \Bigg(
\frac{{\cal{A}}^{\rm 1PI}_{HHh}(m_h^2,M_H^2,M_H^2)}{\l_L v}
-\frac{\delta v}{v} -\frac{1}{2}\delta Z_h -\delta
Z_H \Bigg)_\infty.
\end{equation}

\noi The coefficient of the ultraviolet divergent part is nothing but the one-loop $\beta$ constant ($\beta_{\l_L}$) of $\l_L$. 

\beqn
\label{deltalLMS}
\frac{\delta^{\overline{{\rm MS}}} \l_L}{\l_L} =\beta_{\l_L} C_{{\rm UV}}, \quad C_{{\rm UV}}=\frac{2}{\varepsilon} - \gamma_E+\ln(4\pi),
\eeqn
where $\varepsilon=4-d$ with $d$ being the number of dimensions in dimensional regularisation and $\gamma_E$ being the Euler constant. As discussed at length in Ref.~\cite{Belanger:2017rgu}, a general scheme can be defined as

\beqn
\label{deltalLg}
\frac{\delta \l_L}{\l_L} =\beta_{\l_L} \left(C_{{\rm UV}}+\ln (\bar{\mu}^2/Q^2_\l)\right), 
\eeqn

where $Q_\l$ is an effective scale that depends on the external momenta (hence the subtraction point)  and the internal masses introduced to define the counterterm, and $\bar{\mu}$ is the scale introduced by dimensional reduction. For the $\overline{{\rm MS}}$, scheme $Q_\l=\bar{\mu}$. \\ For $m_h < 2 M_H$, it is difficult to come up with a straightforward OS scheme for $\lambda_L$ (or equivalently $\mu_2$ once the mass counterterms for the extra scalars have been set). We have therefore chosen an $\overline{{\rm MS}}$ scheme for $\l_L$ according to Eq.~\ref{deltalLMS}. A {\it formal} OS extraction that would work for any configuration of $H$ and $h$ masses could use the cross-section that builds up direct detection, namely $Hq\to Hq$ in the limit of zero $Q^2$ transfer, in effect isolating the $H (M_H^2) \to H(M_H^2) h(Q^2 \to 0)$ vertex. But direct detection involves uncertainties through the introduction of  parameters from nuclear matrix elements. Moreover, the $\l_L$ contribution to direct detection can be swamped by the pure gauge contribution (which we discuss later). 

As stated before, the counterterm for $\mu_2$  is directly related to that of $\l_L$:

\beqn
\delta \mu_2^2=\delta M_H^2 -\frac{v^2}{2} \delta \l_L - \l_L v^2 \frac{\delta v}{v}.
\eeqn

\noi For the annihilation processes we will find that a counterterm for $\l_2$ is not necessary, however an $\overline{{\rm MS}}$ definition based on the corresponding $\beta$ function can be derived. This is shown in Appendix~\ref{sec:betal2}.

\section{A high mass IDM benchmark point}
\label{sec:benchmarks}
Previous studies on the collider and astrophysical constraints of the IDM parameter space~\cite{LopezHonorez:2006gr, Agrawal:2008xz, Lundstrom:2008ai, Andreas:2009hj, Arina:2009um, Dolle:2009ft, Nezri:2009jd, Miao:2010rg, Gong:2012ri, Gustafsson:2012aj, Swiezewska:2012eh, Wang:2012zv, Goudelis:2013uca, Krawczyk:2013jta, Osland:2013sla, Abe:2015rja, Arhrib:2015hoa, Blinov:2015qva, Diaz:2015pyv, Ilnicka:2015jba, Belanger:2015kga, Carmona:2015haa, Kanemura:2016sos, Eiteneuer:2017hoh, Ilnicka:2018def, Kalinowski:2018ylg} have delineated two regions with a viable DM candidate that provides the correct relic density of DM. The first one is dubbed the ``low mass regime'' where $M_H \approx M_h/2$ and the second the ``high mass regime'' where $M_H \gtrsim 500$ GeV. To show the impact of a more precise calculation of the annihilation cross-sections that enter the relic density, in this paper we start by finding a point that passes the relic density constraints based on a tree-level calculation. For this we use the code {\tt micrOMEGAs}~\cite{Belanger:2013oya, Belanger:2018mqt}. The relic density constraint set by PLANCK~\cite{Ade:2015xua}, \begin{equation}
\label{omhexp}
 \Omegah = 0.1197 \pm 0.0022,
\end{equation}
is imposed. 

The characteristics of our benchmark point are the following

\beqn
\label{benchmarkheavy}
M_H&=&550 \; {\rm GeV},\quad M_A= 551 \; {\rm GeV},\quad M_H^\pm= 552 \; {\rm GeV}, \nonumber \\
\l_L&= &0.0193 ,   \quad \lambda_2= 0.01  \nonumber \\
(\l_3&=&0.0926, \quad \l_4= -0.0545, \quad \l_5=-0.0181 \; {\rm and} \; \mu_2=549.45 \; {\rm GeV}).
\eeqn
The values between brackets in Eq.~\ref{benchmarkheavy} are derived values. For the SM parameters, we take $M_h=$ 125 GeV, $M_W=$ 80.45 GeV, $M_Z$= 91.19 GeV and $\alpha=1/137$. For these values of the parameters, the calculated relic density (calculated with tree-level cross-sections) is $\Omegah \simeq 0.117$, a value consistent with Eq.~\ref{omhexp}. At tree-level, the cross-section does not depend on $\lambda_2$. Note that the viability of this point relies on the almost degenerate IDM scalar spectrum and the small values of the $\lambda$s. These small values of $\lambda$s automatically ensure that a perturbative calculation can be performed. Moreover, vacuum stability holds~\cite{Branco:2011iw,Goudelis:2013uca} together with the fact that the global minimum is associated with the inert vacuum\cite{Belyaev:2016lok}. The degeneracy in the scalar masses can be viewed as rather fine-tuned~\cite{Goudelis:2013uca}. This degeneracy means that constraints from electroweak precision measurements are easily evaded, in particular, the custodial isospin symmetry parameter $T$~\cite{Kanemura:2016sos}. Indeed in this case, 
\begin{equation}
 \Delta T \simeq \frac{1}{24 \pi^2 \alpha v^2}\left(M_{H^\pm}-M_A\right)\left(M_{H^\pm}-M_H\right),
\end{equation}
is vanishingly small. 

Far more stringent is the constraint from the spin-independent DM-nucleon cross-section for direct detection. In this scenario, the one-loop electroweak gauge contribution to the $H$ nucleon cross-section, $\sigma_{HN}^{(g)}$~\cite{Cirelli:2005uq,Klasen:2013btp}, is almost an order of magnitude larger than the tree-level Higgs exchange contribution triggered by $\l_L$, $\sigma_{HN}^{(\l_L)}$~\cite{Honorez:2010re}. One obtains

\begin{equation}
 \sigma_{HN}^{(\l_L)} = f^2 \frac{\lambda_L^2}{4\pi} \Bigg( \frac{m_N^2}{m_H m_h^2} \Bigg)^2,
\end{equation}
where $m_N$ is the nucleon mass, and $f \sim 1/3$ is the nucleon form factor. With  $M_H \gg M_W$, one can write

\beqn
\mathcal{R}(\lambda_L)&=&\frac{\sigma_{HN}^{(g)}}{\sigma_{HN}^{(\l_L)}} \sim \Bigg(6\pi \frac{\alpha^2}{\l_L s_W^4}\Bigg)^2 \Bigg(\frac{M_H}{8 M_W}\Bigg)^2 \Bigg(1+\frac{M_h^2}{M_W^2}\Bigg)^2 \nonumber \\
&\sim 9& \quad \text{for} \; \l_L=0.019  \; {\rm and} \; M_H=550 \; {\rm GeV}.
\eeqn

This benchmark point passes the present XENON1T~\cite{Aprile:2018dbl} constraint. However, further improvement in the experimental sensitivity from direct detection experiments could make the viability of this benchmark point difficult even if the relic density constraint is passed.  

The degeneracy in the scalar masses means that the relic density is built up by a few co-annihilation channels. They mainly annihilate into vector bosons. The percentage contribution of each channel to the relic density is 
\begin{equation}
\label{list_7proc}
\begin{cases}
H H \to W^+ W^- \quad (18\%), \\
H H \to ZZ  \quad (14\%), \\
H^+ H^- \to W^+ W^- \quad (13\%), \\
A A \to W^+ W^-  \quad (9\%), \\
H^+ H \to W^+ \gamma \quad (8\%),  \\
AA \to ZZ  \quad (7\%), \\
H^+ A \to W^+ \gamma \quad (6\%).
\end{cases}
\end{equation} 
\noi These weights are given by {\tt micrOMEGAs} based on tree-level calculations. The other channels contribute less than or equal to 5\%, including the $hh$ final state channel.  \\
\noi Out of the 7 cross-sections, those with annihilations to a photon,  $H^+ A \to W^+ \gamma$ and $H^+ H \to W^+ \gamma$ are driven solely by gauge couplings and are, at tree-level, independent of $\l_{L,A}$. The other 5 interactions are sensitive to $\l_L$. However, considering the small value of $\l_L$ in our benchmark example, to a good approximation these cross-sections are also dominantly (though not totally) driven by gauge interactions so we can write $\sigma_{HH \to W^+ W^-} \sim \sigma_{AA \to W^+ W^-}  \sim \sigma_{H^+H^- \to W^+ W^-} = 2 c_W^4 \sigma_{HH \to ZZ} = 2 c_W^4 \sigma_{A A \to Z Z}$. The weights quoted in Eq.~\ref{list_7proc} are a measure of the relative importance of the corresponding cross-sections diluted by the the Boltzmann factor. We will now look at the one-loop corrections that affect the 7 dominant processes given in Eq.~\ref{list_7proc} which, at tree-level, contribute more than $5\%$ to the relic density. We will compute these processes for a wide range of velocities.

\section{Annihilation cross-sections at one-loop order}
\label{sec:loopresults}
\subsection{Some important technicalities}
\label{loopcalc-tech}
The calculation of the relic density requires the dependence of the different relevant cross-sections on the relative velocity $v$ of the annihilating particles times $v$, $\sigma_{ij} v_{ij}$, where $i,j$ stand for the annihilating/co-annihilating particles, before applying thermal averaging. Within the standard cosmological model and assuming freeze-out, the latter part is computed quite precisely by {\tt micrOMEGAs}. For two annihilating particles with momenta $p_1$ and $p_2$ and  masses $m_1$ and $m_2$, the relative velocity is defined as 
\beqn
v &=& 2 s \frac{\sqrt{(s - (m_1 + m_2)^2)(s - (m_1 - m_2)^2)}}
{s^2 - (m_1^2 - m_2^2)^2}, \quad s=(p_1+p_2)^2, \nonumber \\
v&=&2 \sqrt{1-4 M_{{\rm DM}}^2/s}=2 \beta \quad {\rm for} \quad m_1=m_2=M_{{\rm DM}}.
\eeqn
However, it is possible to replace the tree-level cross-sections generated by {\tt micrOMEGAs} by corrected cross-sections. This is what we do in order to obtain loop-corrected relic density. In our case, the corrected cross-sections are the one-loop corrected cross-sections for the processes listed in Eq.~\ref{list_7proc}. Most of the steps of the calculation are automated. We rely on {\tt SloopS}~\cite{Baro:2007em, Baro:2008bg, Baro:2009na, Chalons:2012qe} which relies on {\tt LANHEP}~\cite{Semenov:2008jy, Semenov:2014rea} to define the model. {\tt LANHEP} generates the complete set of Feynman rules, applying shifts on fields and parameters and sets the conditions on the generated counterterms in a format compatible with an amplitude generator code. We interface {\tt LANHEP} with  the package bundle {\tt FeynArts}, {\tt FormCalc} and {\tt LoopTools}~\cite{Hahn:1998yk, Hahn:2000kx, Hahn:2000jm}.\\ \noi For each process and for each velocity we check that the virtual corrections (including the counterterms) are ultraviolet finite. To this end we vary the $C_{{\rm UV}}$ (Eq.~\ref{deltalLMS}) parameter by 7 orders of magnitude and check that the result is stable within machine (double) precision. For processes involving charged particles, bremsstrahlung processes $2 \to 2+\gamma$ are generated. The latter is split into two parts, the soft photon radiation and the hard photon radiation. The soft photon radiation for photon energies $E_\gamma <k_c$ with $k_c$ small enough is generated automatically through the factorisation formula which eliminates the one-loop infrared divergence that we regularise with a small finite photon mass. The hard photon radiation is computed numerically. We {\it loop} over a few values of $k_c$ making sure that the soft plus hard part add to a value that is insensitive to $k_c$. This step could be time consuming but we have optimised its automation. When we refer to NLO corrections, we have in mind the full one-loop, the soft and the hard radiation which is of course independent of the regularising photon mass or the intermediate cut-off $k_c$. 

\subsection{Processes at one-loop}
\label{sec:cross-sections}
Our default values for the one-loop corrections are presented in this subsection for a scale $\bar{\mu}=M_H$ taken to define $\l_L$. The scale dependence will be discussed when we convert the results to the level of the relic density calculation.
\subsubsection{$H^+ H^- \to W^+ W^-$}
\begin{figure}[htbp]
\includegraphics[scale=0.7]{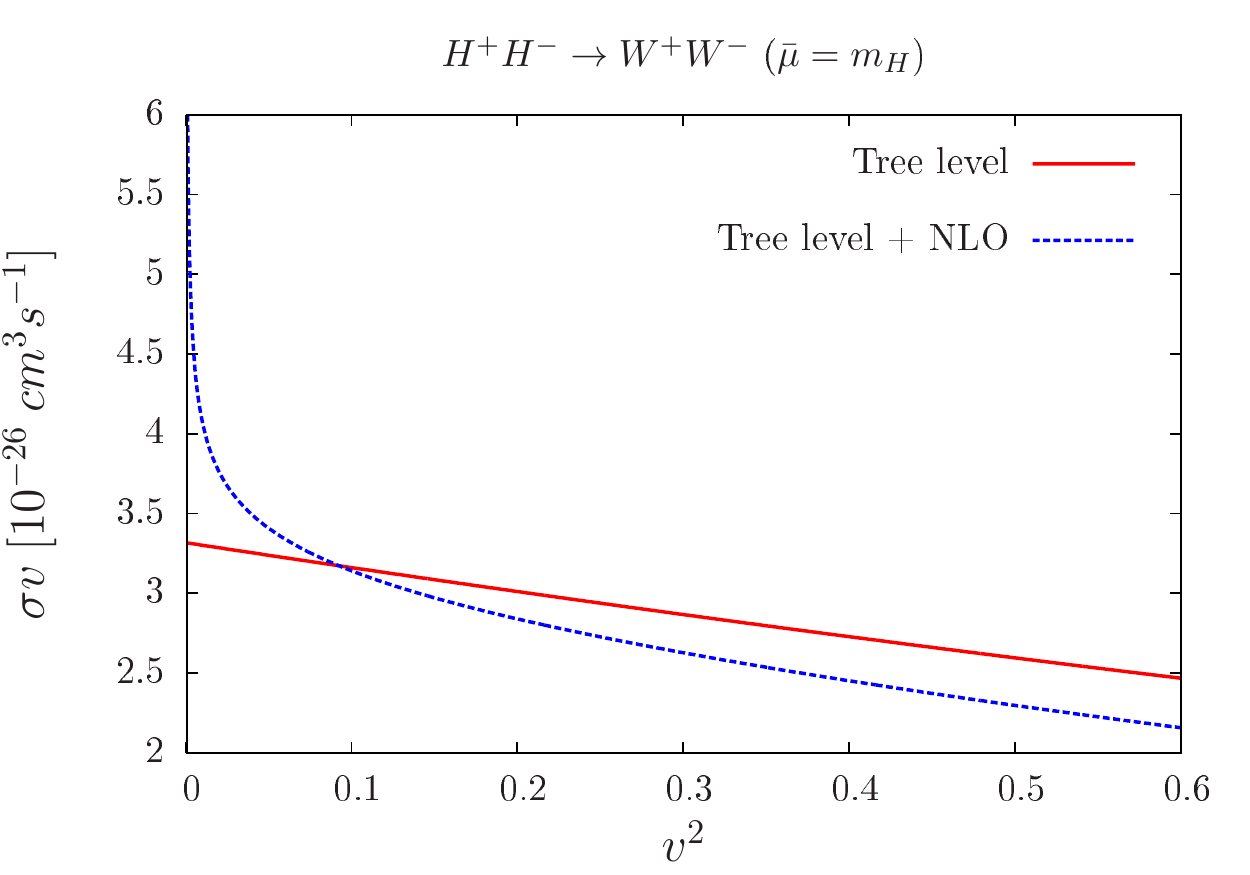}~\includegraphics[scale=0.7]{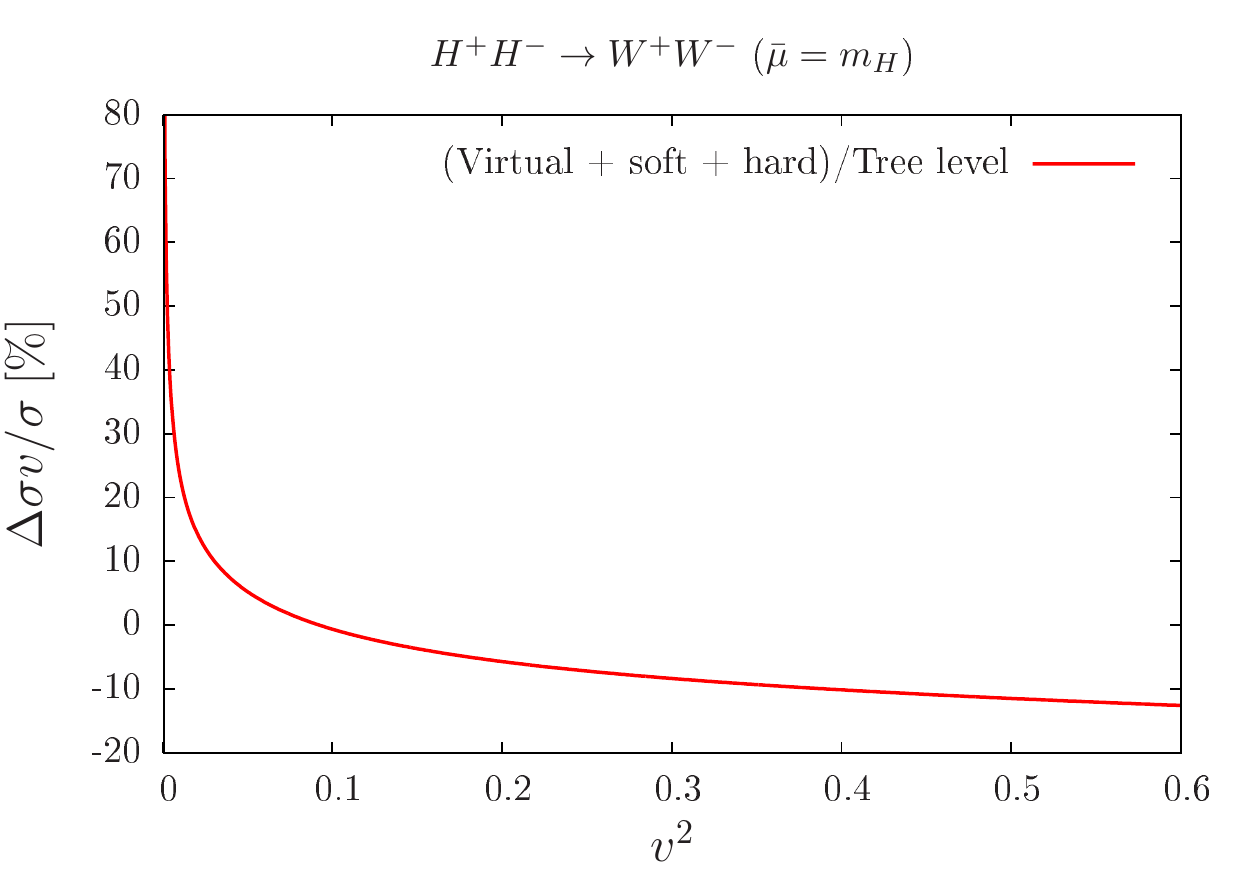}\\
\caption{\it Dependence of the tree-level and one-loop corrected cross-section $H^+ H^- \to W^+ W^-$ with respect to the relative velocity (squared). The right panel gives the percentage correction.}
\label{fig:hphmwpwm}
\end{figure}
We start our discussion with a process whose weight to the relic density (at tree-level) is 13\%. Even though this is not  the most dominant contribution, it helps bring forth an important behaviour. At  tree-level, this process slowly and linearly varies with the relative velocity. This is still the case at one-loop where the full one-loop computation corrects the tree-level result by about $-10\%$ for relative velocities $~\sim 0.2$ and above. However, the one-loop contribution shoots with large positive “corrections” up to extremely low velocities; see Fig.~\ref{fig:hphmwpwm}. This is easily understood as a result of the electromagnetic Sommerfeld effect. The photon exchange between the electrically charged co-annihilating particles at very low relative velocities leads to a relative correction which at one-loop reads as
\beqn
\label{eq:one-loop-sommer}
\frac{\Delta \sigma^{{\rm 1-loop \; Somm.}}v}{\sigma^{{\rm tree}}v}=\frac{\pi \alpha}{v}.
\eeqn
\noi We have checked that our numerical code captures this effect exactly. This one-loop Sommerfeld contribution can be resummed with the result that the tree-level cross-section is turned into

\beqn
\label{eq:sommer_resum}
\sigma^{{\rm resummed}}=S_{{\rm nr}} \; \sigma^{{\rm tree}}, \quad S_{{\rm nr}}=\frac{X_{{\rm nr}}}{1-e^{-X_{\rm nr}}} \quad {\rm and} \; X_{\rm nr}=2 \pi \alpha/v.
\eeqn
\noi Since characteristic velocities for the calculation of the relic density are typically in the range $v \sim 0.2-0.3$, the Sommerfeld enhancement taken either at one-loop or resummed to all orders does not have much of an impact on the relic density. We have checked this feature explicitly. It is however important that our full calculation catches such behaviour at very small velocities quite precisely. While presenting our results for the relic density, this resummation is performed even though its effect is tiny.\\

Note that an electroweak equivalent to the Sommerfeld correction is induced by rescattering through $W$ and $Z$ bosons and even through the SM Higgs boson. For $H^+ H^- \to W^+ W^-$ these low velocity effects are completely swamped by the photon exchange (Sommerfeld effect). For later reference, let us point out that these electroweak equivalents may play a role at very small velocities only if $M_{W,Z,h}/M_{{\rm DM}} \ll 1$, which is not attained in our scenario. The masses of the $W,Z$ and $h$ bosons provide a cut-off to the $1/v$ rise, so the rise of the cross-section, when the massive bosons are involved, is not indefinite. Higgs exchange will be totally negligible considering that the $hHH$ coupling, for example, is controlled by the small $\l_L$.  

\subsubsection{$AA \to ZZ$ and $AA \to W^+W^-$}
\begin{figure}[htbp]
\includegraphics[scale=0.6]{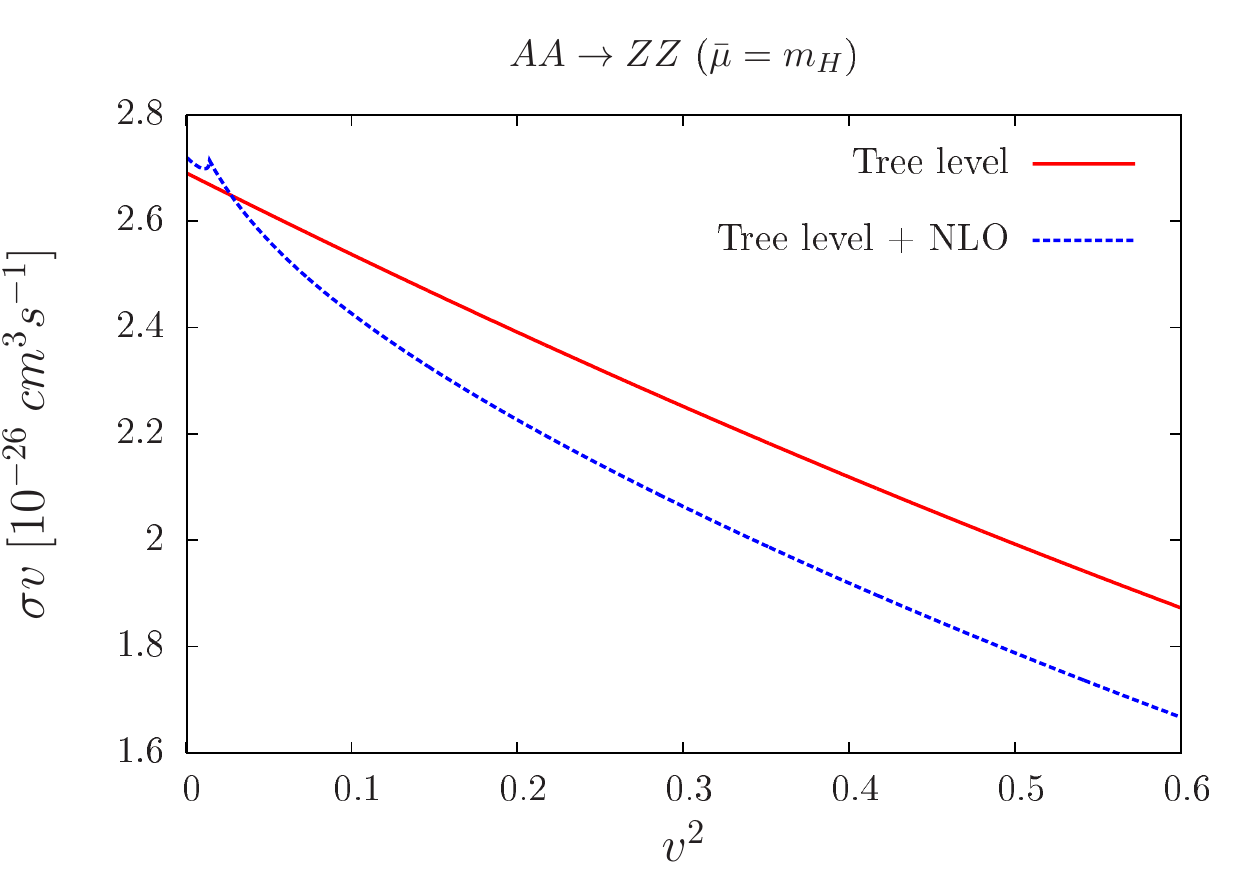}~\includegraphics[scale=0.6]{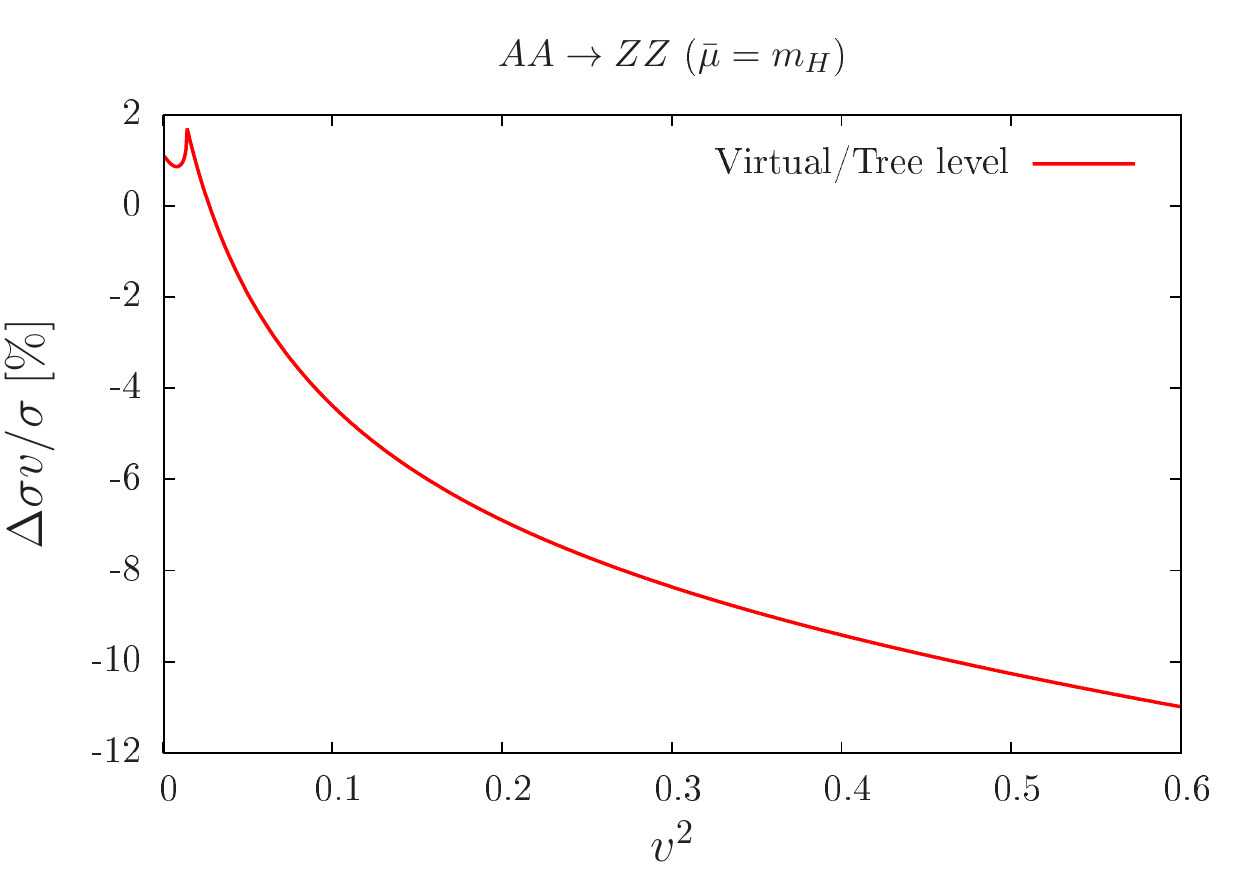}
\includegraphics[scale=0.6]{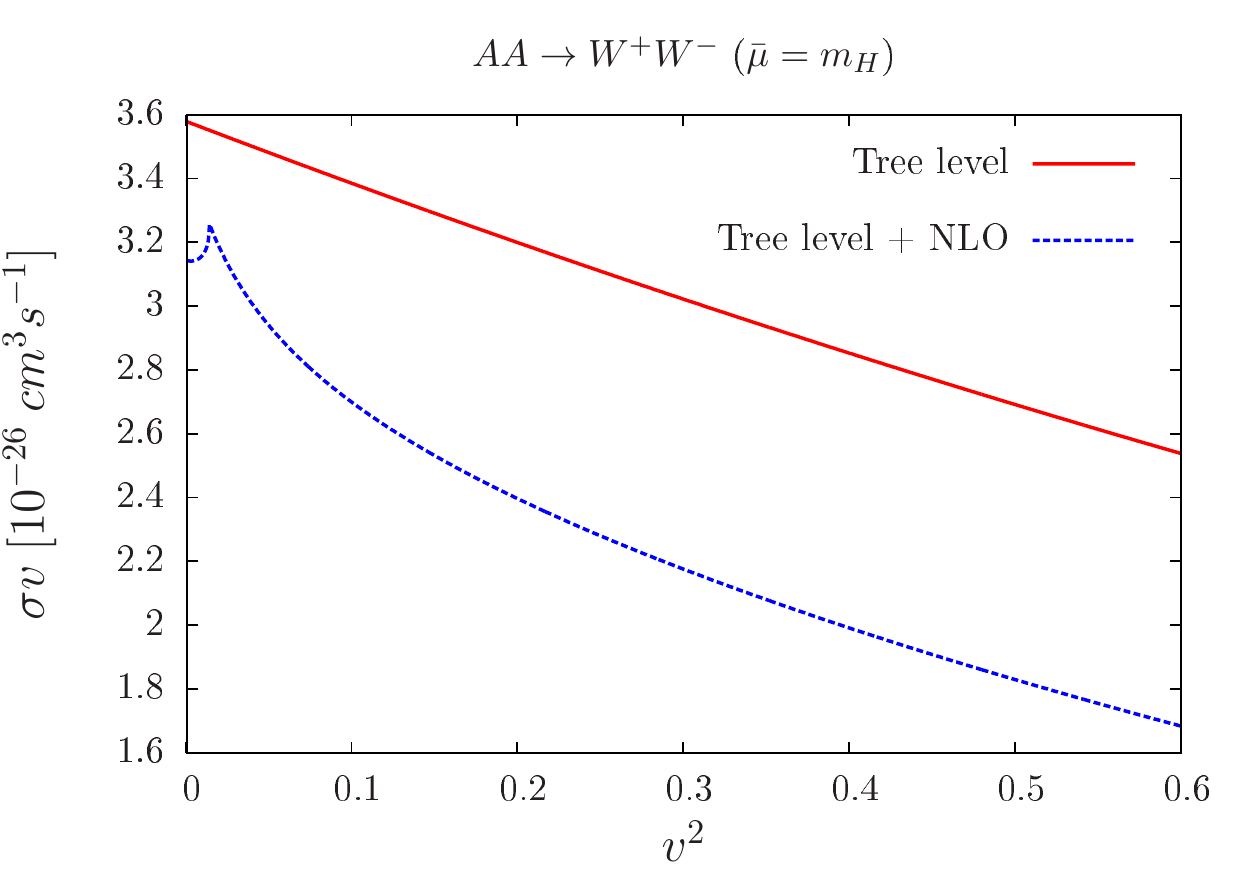}~\includegraphics[scale=0.6]{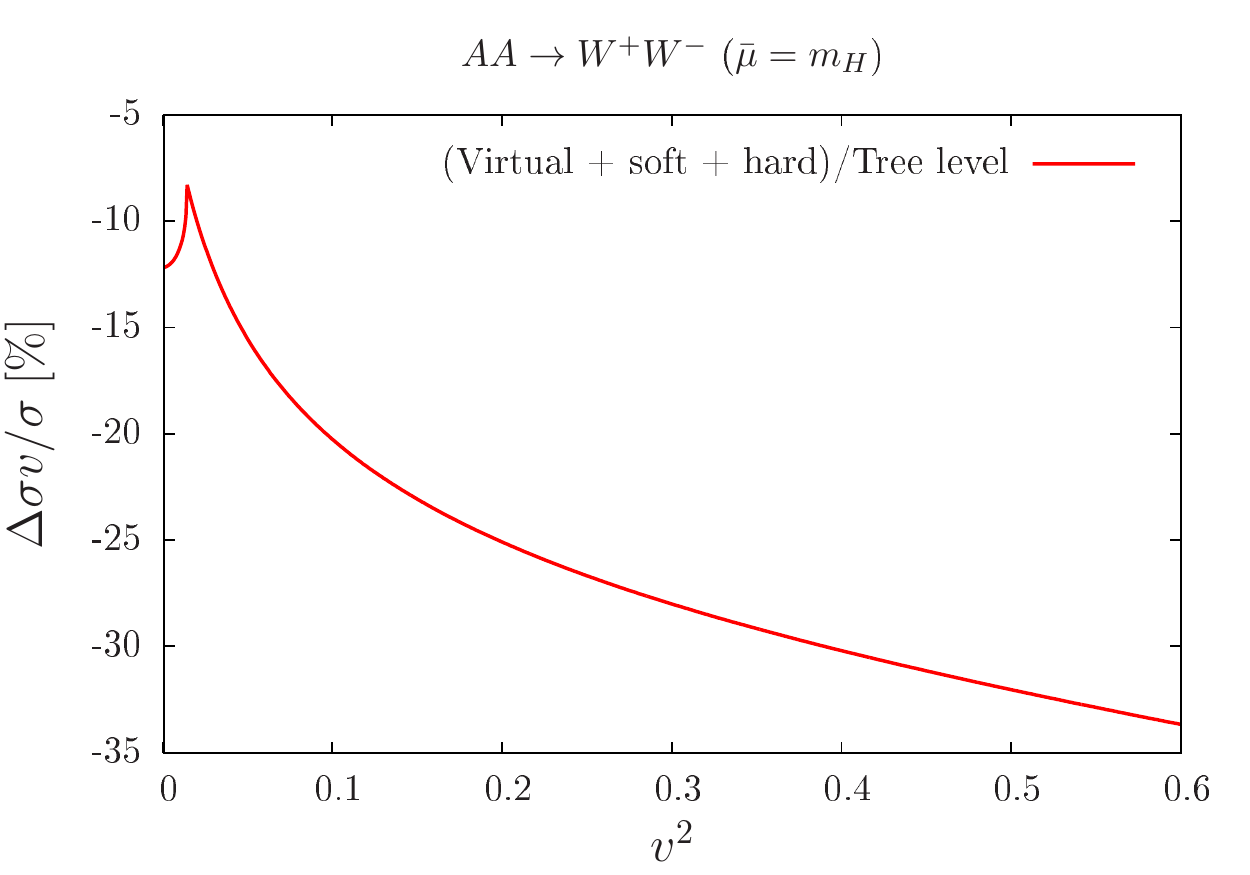} \\ 
\caption{\it Dependence of the tree-level and one-loop corrected cross-section $AA \to ZZ$ (upper panels) and $AA \to W^+ W^-$ with respect to the  relative velocity (squared). The right panels give the percentage correction. }
\label{fig:AAVV}
\end{figure}

$AA \to ZZ$ and $AA \to W^+W^-$ contribute respectively 7\% and 10\% to the relic density at tree-level. The velocity dependence of their cross-sections at tree-level decreases slowly with an almost similar rate. At one-loop, the photon final state radiation affects the $W^+W^-$ channel. The overall corrections are thus larger in the charged channel than in the neutral channel. For $v=0.3$ this correction is a modest $-5\%$ in the $ZZ$ channel but about $-20\%$ in the $W^+W^-$ channel. Nonetheless, the corrections follow a similar trend; see Fig.~\ref{fig:AAVV}. For large velocities, the corrections are largest (and negative) and tend to decrease in absolute values by a contribution that behaves as $1/v^2$ up to $v \sim 0.12$ where it again drops. This behaviour at such small values of the relative velocity is a reminder of the Sommerfeld effect due to the $W$ exchange. In these two processes such effects are only triggered by $W$ exchange (and not by $Z$ exchange ) since no $AAZ$ coupling exists, where $AH^\pm W^\mp$ is operative. This rescaterring, $AA \to H^+ H^-$, also explains why the corrections in the $W^+W^-$ channel are larger, indeed the amplitude for $H^+H^- \to W^+W^-$ is more than twice as large as the $H^+H^- \to ZZ$ counterpart. 


\subsubsection{$HH \to ZZ$ and $HH \to W^+W^-$}
\begin{figure}[htbp]
\includegraphics[scale=0.65]{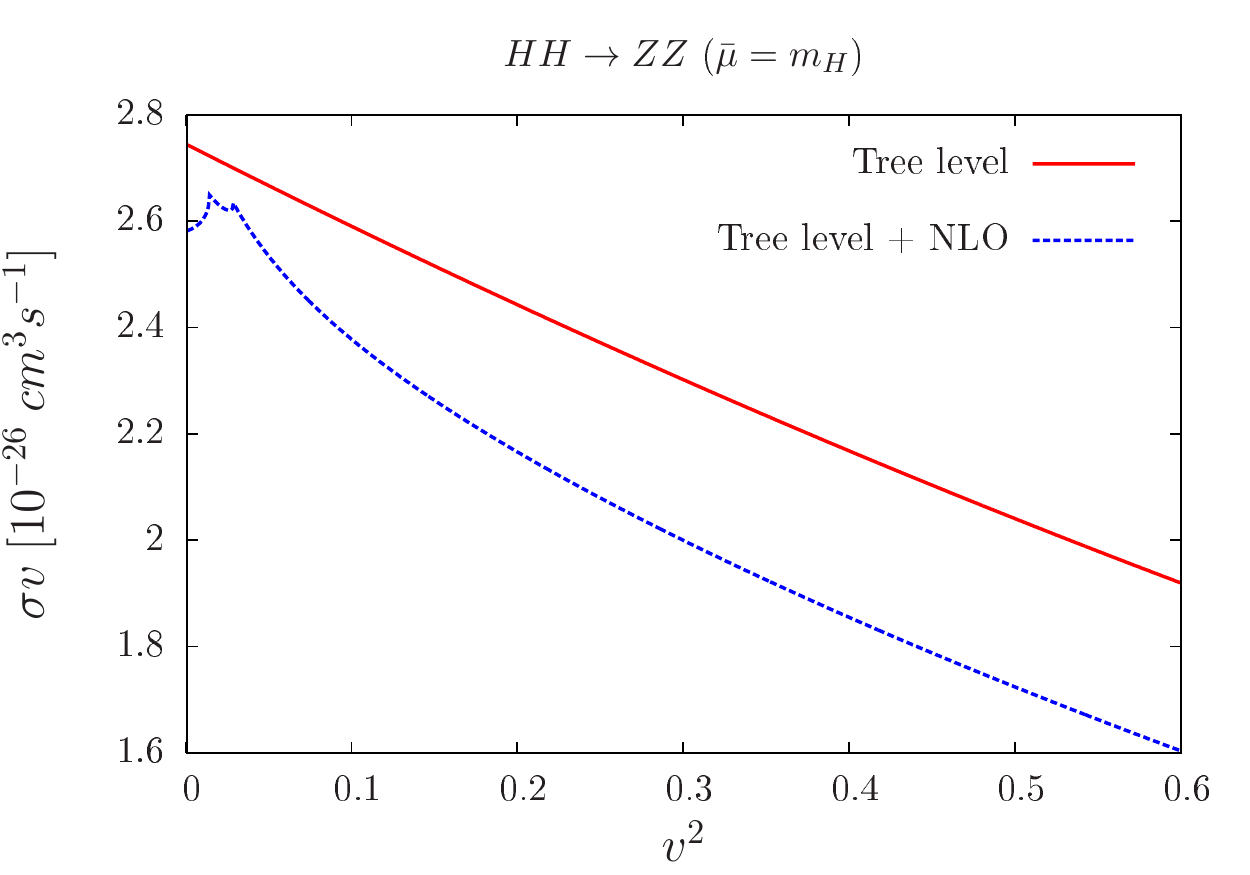}~\includegraphics[scale=0.65]
{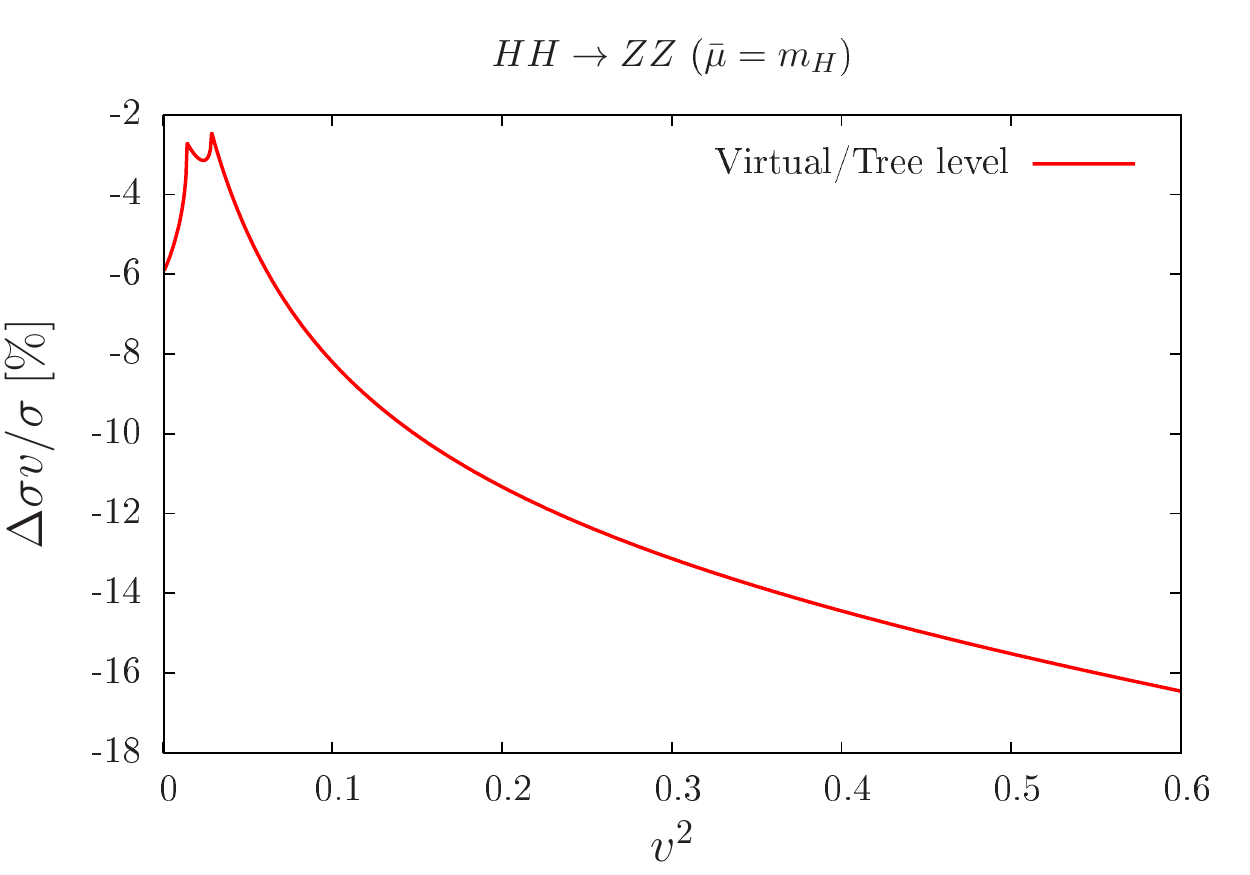}\\
\includegraphics[scale=0.65]{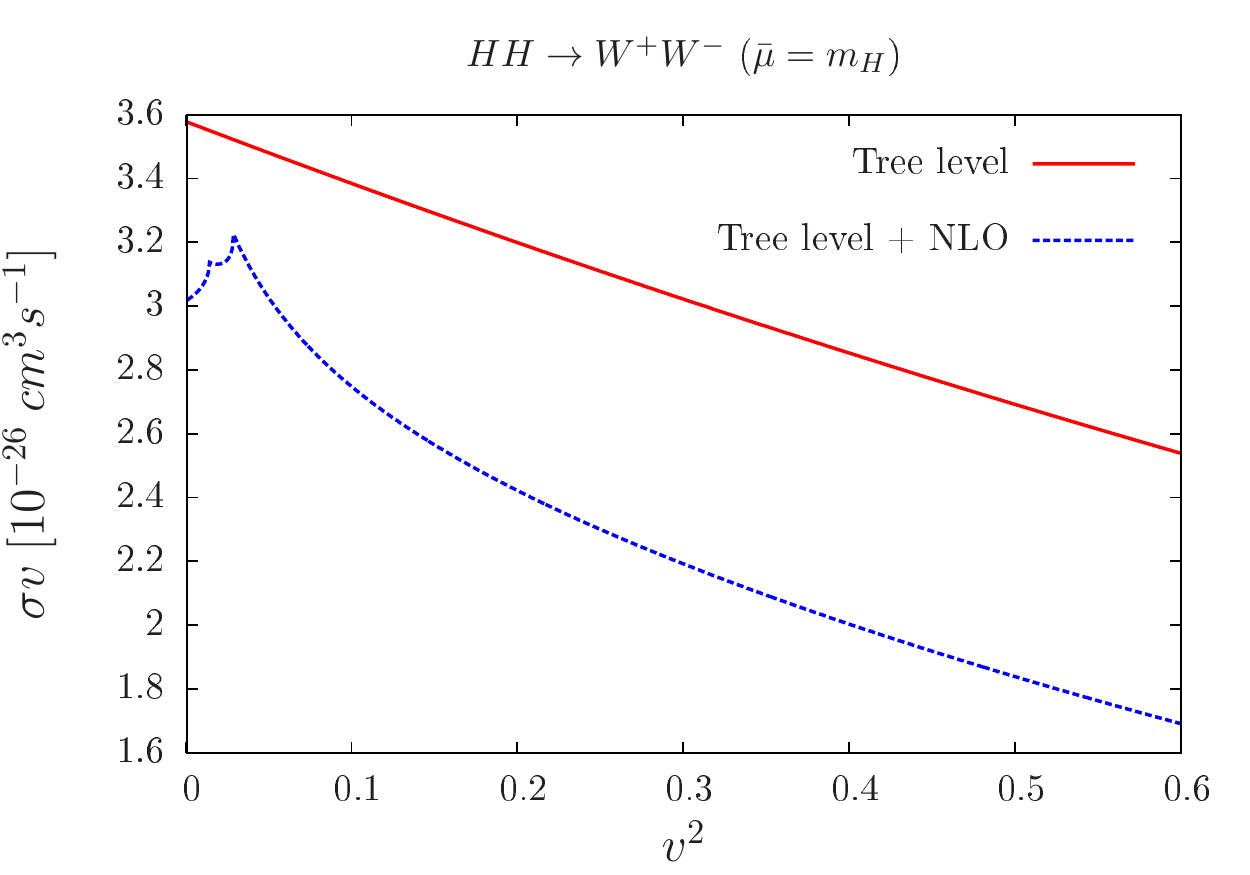}~\includegraphics[scale=0.65]{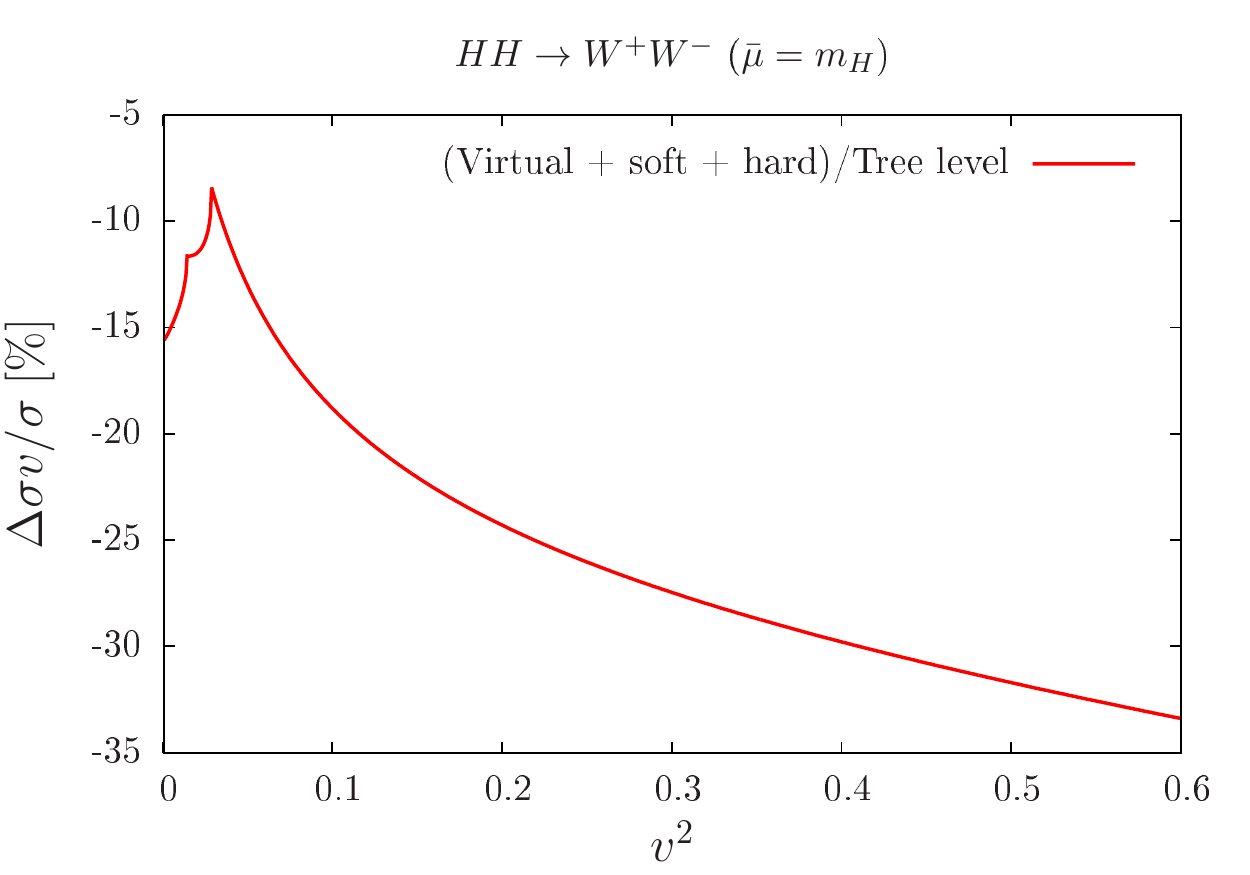}\\
\caption{\it As in Fig.~\ref{fig:AAVV} but for  $H H \to ZZ$ and $HH \to W^+ W^-$. }
\label{fig:HHvv} 
\end{figure}
As expected, for small $\l$s, the $H \leftrightarrow A$  have the same cross-section at tree-level, see Fig.~\ref{fig:HHvv}. Their dependence on velocity is the same, as are the radiative corrections apart from a notable difference for very small velocities. The $HH$ annihilations with respect to the relative velocity now feature two dents, contrary to the $AA$ annihilations where one dent appears. The first of these dents occurs at practically the same location in $v$ as the one that occurs for the $AA$ annihilations. It corresponds to the exchange of the $W$. The second one at slightly larger velocities is due to the $Z$ exchange. Again the corresponding velocities are too small to be relevant for the calculation of the relic density. 

\subsubsection{$H^+ H \to W^+ \g$ and $H^+ A \to W^+ \g$}
\begin{figure}[th!]
\includegraphics[scale=0.65]{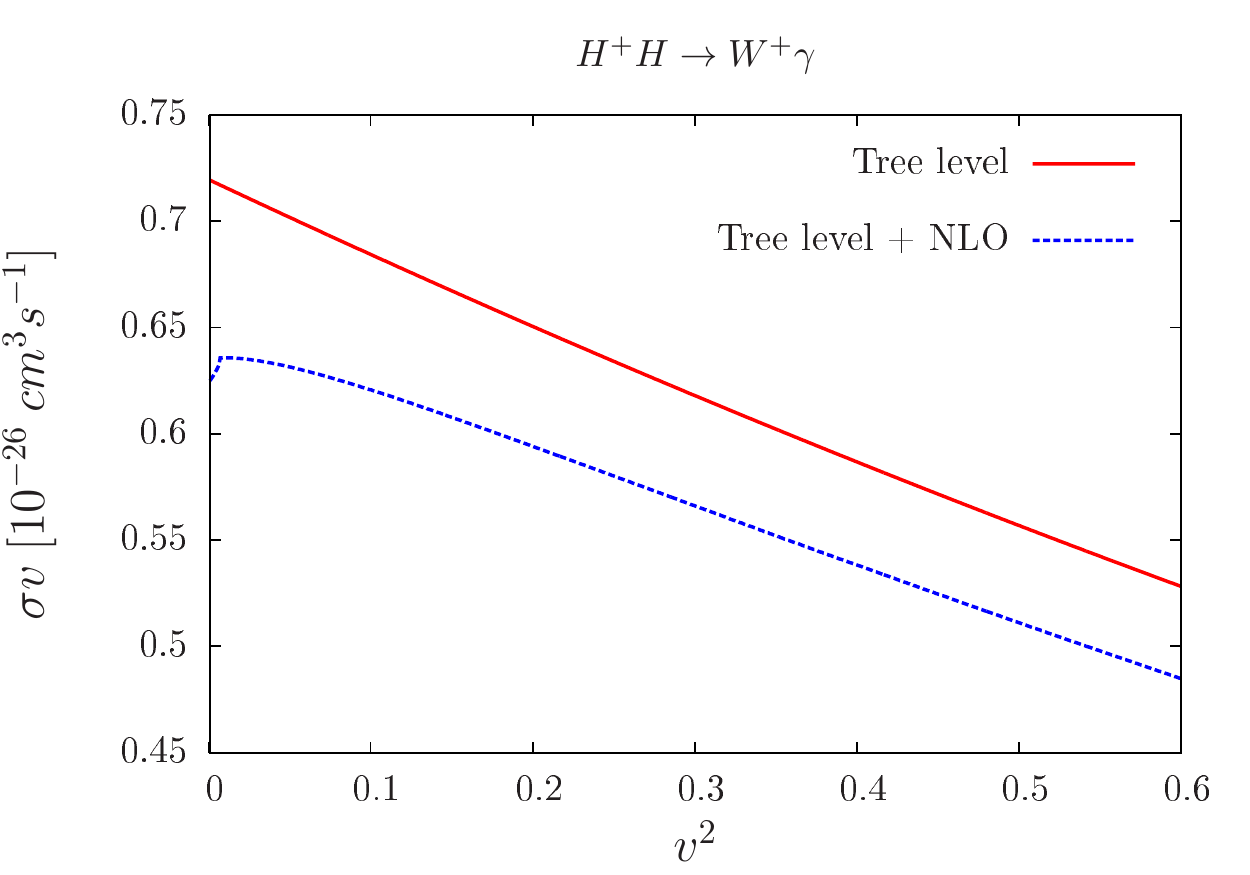}~\includegraphics[scale=0.65]{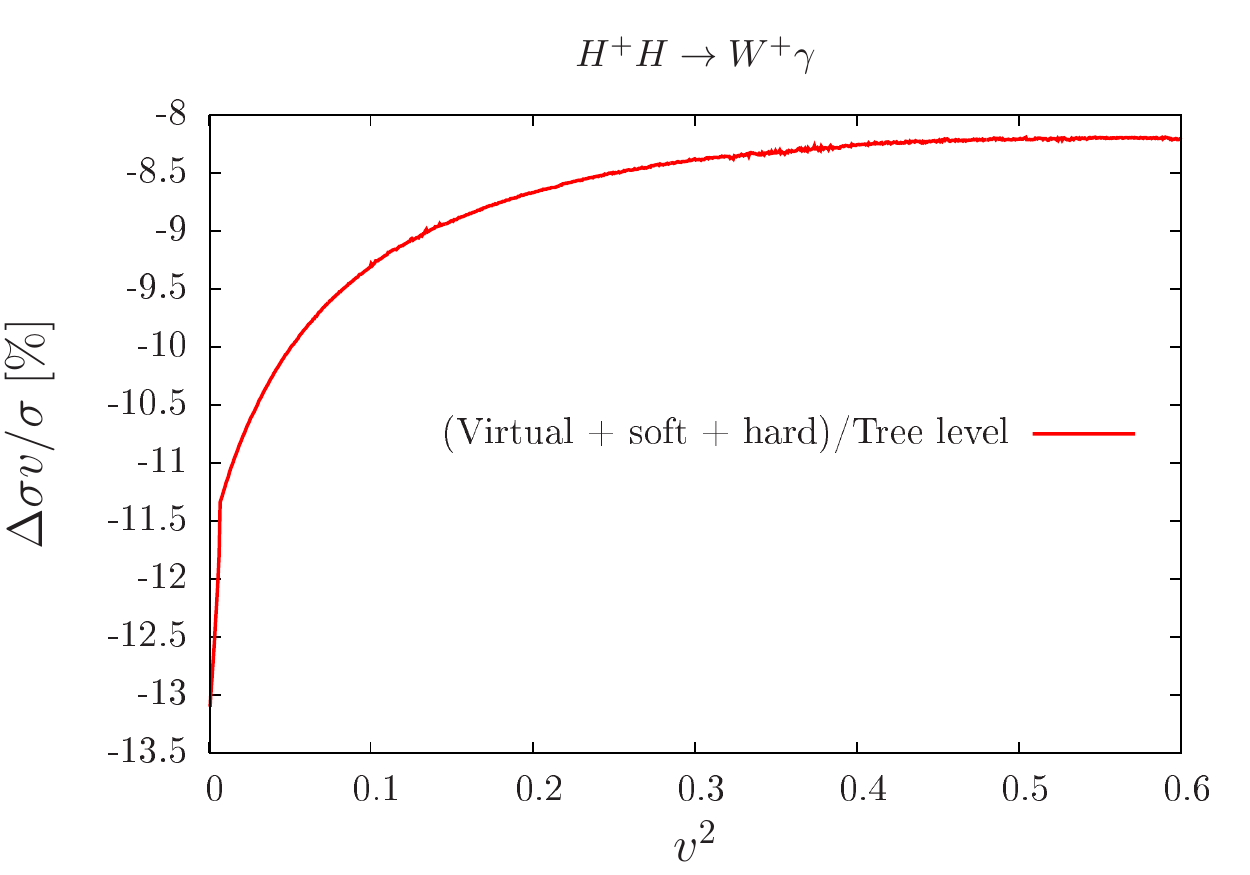}
\includegraphics[scale=0.65]{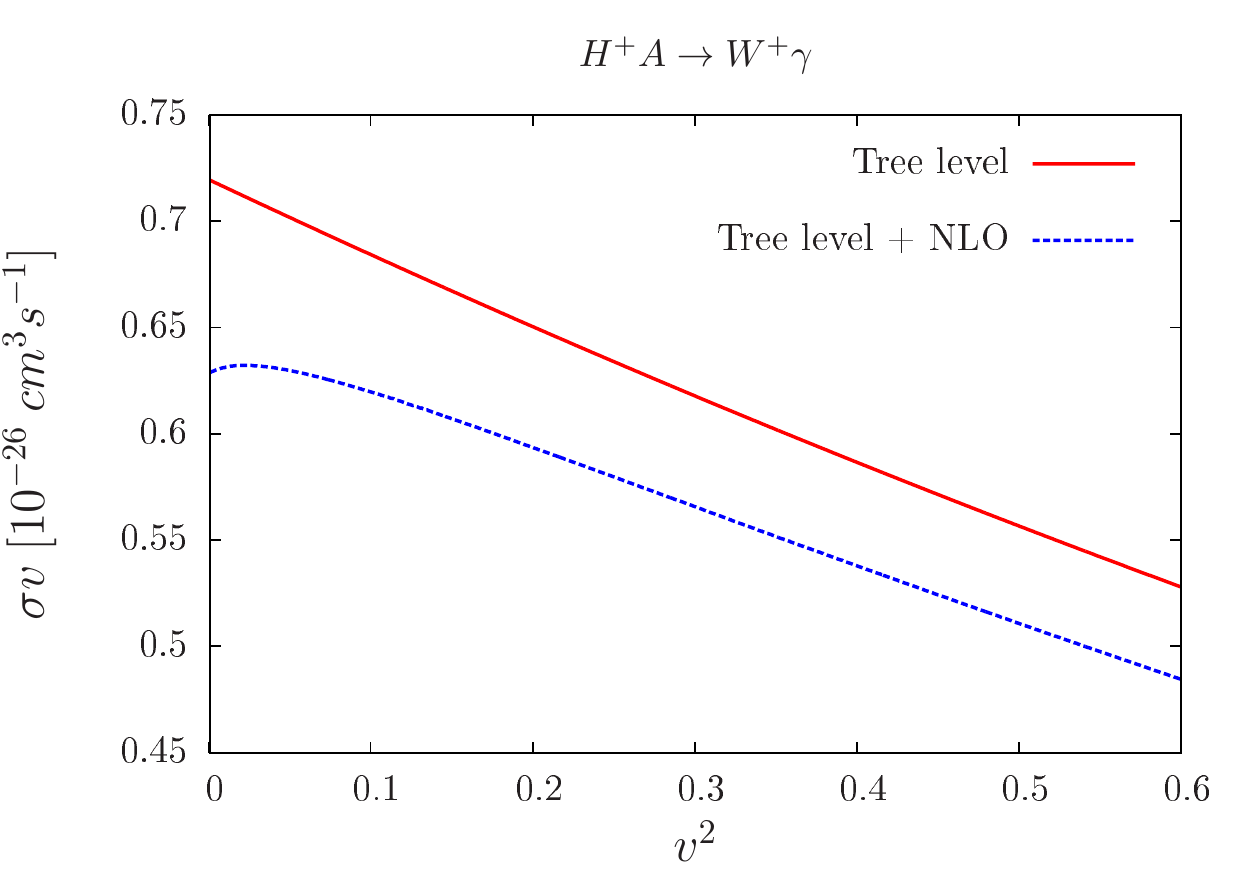}~\includegraphics[scale=0.65]{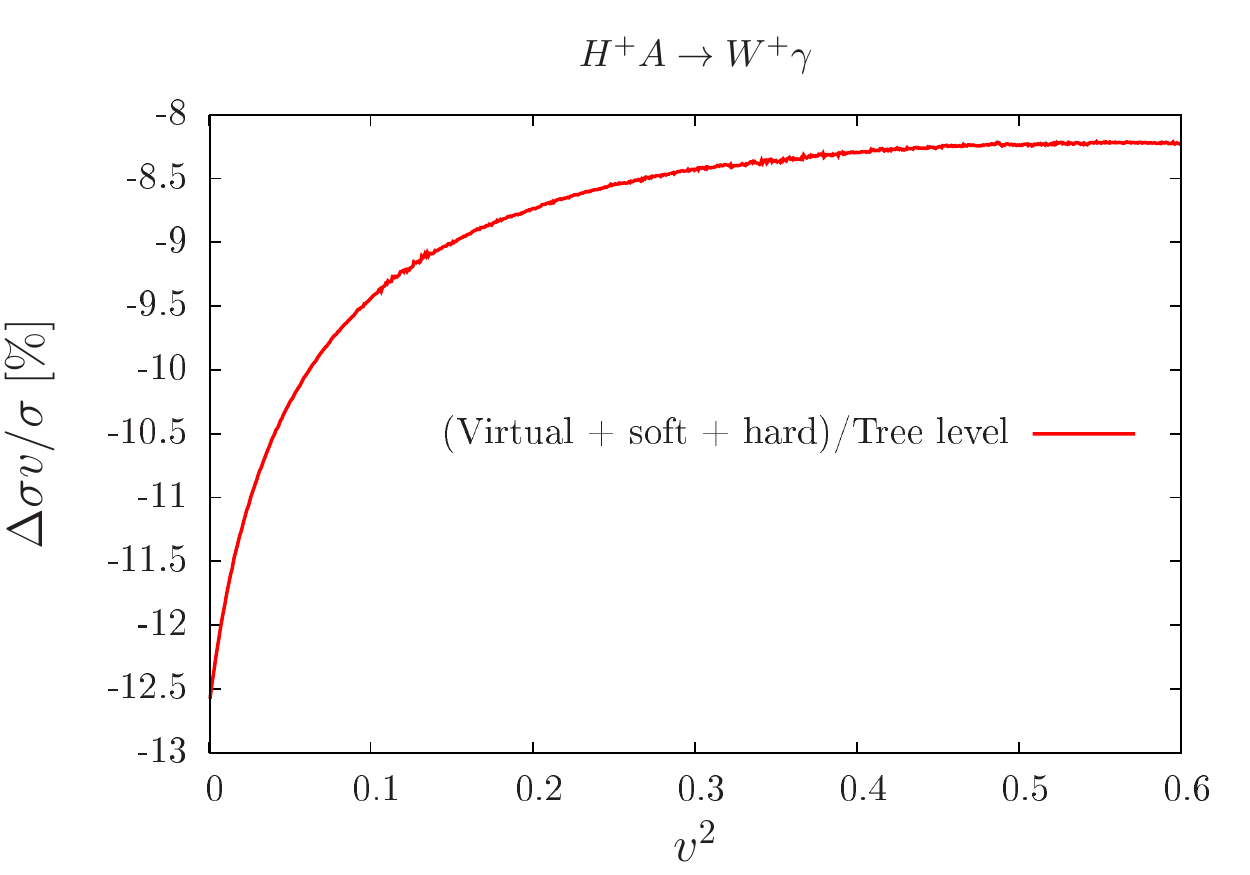}\\
\caption{\it As in Fig.~\ref{fig:HHvv} but for $H^+ H \to W^+ \g$ and $H^+ A \to W^+ \g$ . The one-loop corrections are $\overline{\mu}$-independent since they are totally driven by the gauge interactions.}
\label{fig:HpHwg}
\end{figure}

Again these two cross-sections are practically interchangeable both at the leading and at next-to-leading order. The relative one-loop corrections decrease as the relative velocity decreases, with a correction of about $-10\%$ for a typical velocity, $v\sim 0.3$, see Fig.~\ref{fig:HpHwg}. Since at tree-level, these processes do not depend on $\l_{L,A}$, a fully  on-shell renormalisation is possible with the results that these one-loop cross-sections are  $\overline{\mu}$-independent. 

\section{Relic abundance computations}
\label{sec:relic-one-loop}
Having performed the full one-loop corrections to the 7 processes that make up about $70\%$ of the total relic density at tree-level, we have interfaced our calculations with {\tt micrOMEGAs} by providing the tables for these cross-sections (with the velocity dependence) in lieu of their tree-level value to {\tt micrOMEGAs} for the calculation of the relic density (thermal averaging, freeze-out). The remaining processes ( $H^+ H^- \to \gamma \gamma$, $H^+ H^- \to \gamma Z$, $H^+ H \to Z W^+$, $H^+ A \to Z W^+$, $H^+ H^- \to Z Z$ and $H^+ H^- \to h h$) were kept at their tree-level values. For the loop corrected $H^+ H^- \to W^+W^-$ we compared the result of the relic density with full one-loop calculation, the subtraction of the Sommerfeld correction (Eq.~\ref{eq:one-loop-sommer}), and the replacement of the Sommerfeld contribution with its resummed classical result (Eq.~\ref{eq:sommer_resum}). As expected, we found no noticeable change between the different implementation of the QED Sommerfeld effect. As we saw above, all the cross-sections we calculated at one-loop are affected by a negative correction for $\bar \mu=M_H$. Here we recall that we had taken as input $\alpha=\alpha(0)$, the sign and size of the corrections are not due to the running of $\alpha$. 
Smaller one-loop cross-sections compared to tree-level translate to a larger relic density than derived from tree-level cross-sections. This is corroborated by the value $\Omega h^2=0.12494$ that we find. For $\bar{\mu}=M_H/2$, the loop corrections are smaller, and this naturally translates into a smaller correction to the relic density. Indeed we find a correction of only $\sim -0.3\%$. This would tend to suggest that this scale, for the aforementioned choice of parameters, is optimal for reducing the size of the radiative corrections. 
For $\mu=2 M_H$, the correction to the relic density amounts to $\sim 15.3\%$. The scale variation is large compared to the experimental precision on the relic density. Considering that these scenarios are quite fine-tuned, for an almost degenerate scalar mass spectrum, our calculations show that if one is performing a tree-level analysis one should not strictly impose the very constrained experimental bound on the relic density but one should allow a theoretical uncertainty of at least about 10\% for such benchmark scenarios. We may ask whether the radiative corrections change much the relative weights of the different processes. Table~\ref{table:rel-cont} shows that these relative contributions experience little change, independent of the scale chosen. The most important change is therefore an almost uniform correction on all the cross-sections.

\small
\begin{table}[htp]
\begin{center}
\begin{tabular}{||c|c|c|c|c||}
\hline
Process                      &  LO   & $\mu = m_X$ & $\mu = m_X/2$ &  $\mu = 2 \times m_X$ \\
\hline
$H H \to W^+ W^-$            &  18\% & 16\%        &  18\%         &  14\% \\
$H H \to Z Z$                &  14\% & 14\%        &  14\%         &  14\% \\
$H^+ H^- \to W^+ W^-$        &  13\% & 15\%        &  15\%         &  15\% \\
$A A \to W^+ W^-$            &  9\%  & 8\%         &  9\%          &  7\% \\
$H^+ H \to W^+ \gamma$       &  8\%  & 7\%         &  7\%          &  8\% \\
$A A \to Z Z$                &  7\%  & 7\%         &  7\%          &  8\% \\
 $H^+ A \to W^+ \gamma$       &  6\%  & 6\%         &  6\%          &  7\% \\
$\bullet H^+ H^- \to \gamma \gamma$  &  5\%  & 5\%         &  5\%          &  6\% \\
$\bullet H^+ H^- \to \gamma Z$       &  4\%  & 5\%         &  4\%          &  5\% \\
$\bullet H^+ H \to Z W^+$            &  3\%  & 3\%         &  3\%          &  3\% \\
$\bullet H^+ A \to Z W^+$            &  3\%  & 3\%         &  2\%          &  3\% \\
$\bullet H^+ H^- \to Z Z$            &  2\%  & 2\%         &  2\%          &  2\% \\
 \hline
\end{tabular}
\end{center}
 \caption{\it Relative contributions to the relic abundance with and without corrections. Note that although the cross-sections for the last 5 processes (identified with $\bullet$) are not loop corrected, their relative contribution could change.}
\label{table:rel-cont}
\end{table}

\newpage
\section{News from the dark sector: Impact of $\l_2$}
$\l_2$ does not enter the calculation of the annihilation cross-sections at tree-level. But, just as the $Z$ decay to muons does not depend on the top quark mass at tree-level, at one-loop the top quark makes its effect felt. For the case at hand, there is less subtlety for the non-decoupling of the self-coupling in the dark sector. $HH$ can rescatter before annihilating to gauge bosons. The rescattering $HH \to HH$ involves the self-coupling $\l_2$. One therefore expects the one-loop annihilating cross-sections to depend on $\l_2$. To investigate this effect, we retain the same value of $\l_L$ and consider two other values of $\l_2$, both well within the perturbative and positivity of the potential bound, $\l_2=0.1$ and $\l_2=1$. Our results are shown in Table~\ref{tab:rel-contl2}.
\begin{table}[hb!]
\label{table:rel-contl2}
\begin{center}
\begin{tabular}{||c|c|c|c||}
\hline
$\l_2$ & $\mu=M_H$         & $\mu=M_H/2$       & $\mu=2 M_H$\\
\hline
0.01  & 0.12494 (6.9\%)    & 0.11652 (-0.3\%)  & 0.13469 (15.3\%)\\
0.1   & 0.12210 (4.5\%)    & 0.11843 (1.3\%)   & 0.12601 (7.8\%)\\
1     & 0.09950 (-14.9\%) & 0.14163 (21.2\%)  & 0.07683 (-34.3\%)\\
\hline
\end{tabular}
\end{center}
\caption{\it Dependence of the relic density on the parameter $\l_2$ and the influence of the scale variation. The percentage change is shown within brackets.}
\label{tab:rel-contl2}
\end{table}
We observe that there is a noticeable albeit small change when $\l_2$ is increased to $0.1$, the scale variation is reduced and the total electroweak corrections to the relic density are below $7.8\%$. The case $\l_2=1$ is much more interesting. The corrections are now quite large for each of the three renormalisation scales $M_H/2,M_H$ and $2M_H$. For all of these three scales, the tree-level benchmark point would be ruled out. However, we note that the large scale uncertainty with corrections ranging between +21.2\% for $\mu=M_h/2$ and -34.3\% for $\mu=2 M_H$ means that a judicious scale choice, within the range $M_H/2$ to $2M_H$, can minimise the corrections. A more thorough one-loop analysis is in order by studying other scenarios with a larger range of values for the other quartic couplings. One could find points not allowed by a tree-level analysis that could be  validated by a one-loop analysis. We leave this interesting analysis for a future publication. It looks however that although the virtual effect of $\l_2$ is not at all negligible it (fortunately or unfortunately) introduces also a non-negligible scale variation to the corrections. More importantly, compared to a tree-level treatment, one loop corrections introduce not only a scale uncertainty which is manageable for small values of $\l_2$ but also a parametric dependence (dependence on $\l_2$) which is not caught by a tree-level treatment. In fact, this $\l_2$ dependence turns out to be even larger than the scale dependence. One could in fact use the measurement of the relic density to constrain $\l_2$. 

\section{Conclusions}
\label{sec:conclusions}
The experimental value of the relic density as extracted from PLANCK data is now at the per-cent level. For many particle physics models of dark matter this is a very stringent bound that reduces drastically the range of the parameters in that model. Assuming a standard cosmological model based on freeze-out, the restriction on the parameter space of the model arises from the contribution of the annihilation and co-annihilation cross-sections that build up the evaluation of the relic density. Unfortunately, most analyses are based on tree-level evaluations of these cross-sections. The level of precision on the experimental bound on the relic density calls for a theoretical prediction that should go beyond a tree-level evaluation of these cross-sections. Such a programme has been set up for the minimal supersymmetric model~\cite{Baro:2007em, Baro:2008bg, Baro:2009na, Boudjema:2011ig, Chalons:2012qe, Boudjema:2014gza, Han:2018gej} and the next-to-minimal supersymmetric model~\cite{Belanger:2017rgu}. After a first exploratory investigation in Ref.~\cite{Banerjee:2016vrp}, the present paper extends this programme to the IDM. As such this paper presents a full systematic renormalisation of the model and specialises in the first application into the so called heavy mass scenario. In fact, in order to be fully perturbative, here we have covered a scenario with heavy scalar masses not heavier than 550 GeV. We performed full one loop calculations to 7 annihilation/co-annihilation processes. We interfaced these corrected cross-sections with {\tt micrOMEGAs} to turn these cross-sections into a more precise evaluation of the relic density assuming the standard freeze-out mechanism. The one-loop calculations are implemented in an automated code for loop calculations. We have used a mixed scheme where most of the parameters are defined on-shell, based on the physical masses of the model. Having exhausted all masses in the model to fully define the model, one parameter is defined in $\overline{{\rm MS}}$, the coupling of the scalar DM, $H$, to the SM Higgs. We find that the one-loop corrections to the relic density for this particular mass vary from $-34\%$ to $+15\%$ depending on the renormalisation scale chosen to define the $h-H-H$ coupling and most importantly on the value of  the coupling $\l_2$ which measures the interaction solely within the dark sector between the extra scalars. This is an indirect effect that should be taken into account especially for $\l_2$ of order $1$. Its effects can be larger than the renormalisation scale uncertainty of the one-loop calculation, else the relic density calculation can be used to set a limit on the dark-sector interaction. A tree-level calculation of the relic density is totally insensitive to these couplings that describe interaction within the dark sector. Preliminary investigations for DM masses beyond 750 GeV shows that electroweak Sommerfeld effects become important and that some resummation needs to be performed and merged with perturbative purely one-loop effects like those triggered by rescattering in the dark sector (the indirect effects of $\l_2$). We leave the study of this mass range to a forthcoming publication. Note that a for masses beyond 1 TeV, a purely non perturbative treatment has been given in~\cite{Hambye:2009pw, Biondini:2017ufr}. Other viable parameter space of the IDM is the low mass regime with $M_H \approx M_h/2$. This requires the calculation of $2 \to 3$ processes at one-loop for the evaluation of the relic density. We also leave this application for a forthcoming publication. 

\appendix

\section{Appendix: Feynman rules}
\label{sec:appendix}
We give below the Feynman rules of the tri-linear and quadri-linear couplings among the scalars of the model using the
parametrisation of  Eq.~\ref{inputparam}.
\subsection{Cubic Higgs couplings}
\begin{align}
 h-h-h &: -3i\left(\frac{M_h^2}{v}- \frac{T}{v^2}\right), \\
\label{Hhhcoup}
 h-H-H &: -2i \left(\frac{M_H^2-\mu_2^2}{v}\right) = - i \l_L v, \\
 h-A-A &: -2i \left(\frac{M_A^2-\mu_2^2}{v}\right) = - i \l_A v, \\
 h-H^+-H^- &:  -2i \left(\frac{M_{H^\pm}^2-\mu_2^2}{v}\right)= - i \l_3 v.
\end{align}

\subsection{Quartic Higgs couplings}
\begin{align}
 h-h-h-h  &:  -3i\left(\frac{M_h^2}{v^2}- \frac{T}{v^3}\right), \\
 h-h-H-H &: -2i \left(\frac{M_H^2-\mu_2^2}{v^2}\right) = - i \l_L,\\
 h-h-A-A &: -2i \left(\frac{M_A^2-\mu_2^2}{v^2}\right) =- i \l_A,\\
 h-h-H^+-H^- &:  -2i \left(\frac{M_{H^\pm}^2-\mu_2^2}{v^2}\right)= - i \l_3, \\
 H-H-H-H &: - 6 i \l_2, \\
 A-A-A-A &: - 6 i \l_2, \\
 H-H-A-A &: -2 i \l_2, \\
 H-H-H^+-H^- &: -2 i\l_2, \\
 A-A-H^+-H^- &: -2 i\l_2, \\
 H^+-H^--H^+-H^- &: -4 i \l_2.
\end{align}

\section{Counterterm for $\l_2$}
\label{sec:betal2}
   The $\msbar$ counterterm for $\l_2$ can be obtained from adapting the beta functions of the two Higgs doublet model to the IDM given in Ref.~\cite{Chowdhury:2015yja} for example. In particular,  we can write

\begin{equation}
\label{delta_l2_ms}
 \delta \l_2^{\msbar} = \frac{1}{32\pi^2}\left(\hat{\beta}_{\l_2}^S +
\hat{\beta}_{\l_2}^g \right) C_{{\rm UV}},
\end{equation}

with 
\begin{align}
 \hat{\beta}_{\l_2}^S &= 24 \l_2^2 + 2 \l_3^2 + 2 \l_3 \l_4 + \l_4^2 + \l_5^2, \\
 \hat{\beta}_{\l_2}^g &= \frac{3}{8}\Bigg(3g^4 + g^{'4}+ 2 g^2g^{'2} - 3 \l_2
\left(3 g^2 + g^{'2}\right)\Bigg),
\end{align}\noi 
where $g = e/s_W, g'=e/c_W$.

\acknowledgments
We thank Alexander Pukhov for several helpful discussions at various stages of this work. The work of SB is supported by a Durham Junior Research Fellowship CO-FUNDed by Durham University and the European Union, under grant agreement number 609412. NC acknowledges financial support from National Center for Theoretical Sciences and also thanks LAPTh for hospitality during the formative stages of the project. HS is supported by the National Natural Science Foundation of China (Grant No.11675033) and by the Fundamental Research Funds for the Central Universities (Grant No. DUT18LK27). This work was initiated within CNRS LIA (Laboratoire International Associ\'e) THEP (Theoretical High Energy Physics) and the INFRE-HEPNET (IndoFrench Network on High Energy Physics) of CEFIPRA/IFCPAR (Indo-French Centre for the Promotion of Advanced Research). SB and HS also thank LAPTh for hospitality where a part of this work was carried out. SB and FB acknowledge the Les Houches workshop series ``Physics at TeV colliders'' 2019 where this work was finalised.

\bibliography{NLO_IDM}{}
\bibliographystyle{unsrt}

\end{document}